\documentstyle[12pt,twoside,cosmion,epsf]{article}
\heads{Physical Grounds for Inflation, Evolution of the Universe, Baryosynthesis, Dark Matter and New Particles}{
Mass function of gravitationally bounded objects in 
the inhomogeneous Universe\ \ {\rm N.A. Arhipova et al.}}

\begin{document}

\Arthead{1}{1}

\Title{Mass function of gravitationally bounded objects in 
the inhomogeneous Universe}
{N.A. Arhipova$^1$, V.N. Lukash$^1$, B.V. Komberg$^1$, E.V. Mikheeva$^1$}
{$^1$Astro Space Center of Lebedev Physical Institute 84/32 Profsoyuznaya st.,
Moscow, 117810, Russia}

%

\Abstract{%
We modify the Press-Schechter formalism to calculate the mass function 
of gravitationally bounded objects in a local Universe with size
$L> 10h^{-1}$Mpc, i.e. separately in voids ($\Omega_{m,L}<\Omega_m$) and 
superclusters ($\Omega_{m,L}>\Omega_m$). 
The dependence of the abundance of gravitationally bounded objects on local
matter density is analyzed. 
}


\section{Introduction}
Although the Universe is spatially flat and homogeneous at large scales
($>300\; h^{-1}$ Mpc), at smaller scales it can be parted in over- and 
under- density regions. We will label these regions as superclusters and 
voids, respectively. The correctness of such classification is justified by 
a formal smoothing of the matter density field with a window of large size 
$L>10h^{-1}$Mpc, the obtained smoothed density is then characterized 
by the local density parameter $\Omega_{m,L}$: by definition, the regions
with $\Omega_{m,L}{}^>_<\Omega_m$ are superclusters and voids, respectively.
Obviously, the 3D volume of the Universe is separated on these two 
regions (depending on the formal parameter $L$) which can be refered as
the most extended 'elements' of LSS. Observationally, a typical difference
$\Delta\equiv\Omega_{m,L}-\Omega_m$ is about $\sim 0.1\div 0.2$
in largest voids and superclusters, $L\sim 100h^{-1}$Mpc. This scale is 
physically related with the cosmological horizon at the equipartition epoch.   

The difference between superclusters and voids appears in many features. In
addition to the different sign of $(\delta\rho)_L/\rho$ (averaged with the 
scale of the consequent element of LSS), superclusters and voids can be 
distinguished by their population. In supercluster we observe the brightest 
galaxies, their groups and clusters, galaxies with AGNs, and others, whereas 
all these bright objects avoid voids (Dalcanton\cite{1}, Mobasher \& 
Trenthom\cite{9}). The current observational data demonstrate that the main 
population of voids are dwarf galaxies (Shade\cite{14}, Pustilnik et 
al\cite{11}, Linder \& Einasto\cite{7}) and, may be, $Ly_{\alpha}$-clouds 
(Rees\cite{13}, Rauch\cite{12}).
This difference in the populations have to be appear in mass function, but 
these researches are at the beginning now. We try to solve this problem by 
modification of Press-Schechter formalism (Press \& Schechter\cite{10}) which 
allows to calculate mass functions in voids and superclusters separately, in 
other words in the inhomogeneous Universe.

\section{Press-Schechter formalism in the inhomogeneous Universe}

The Press-Schechter (P\&S) formalism is based on two main assumptions:
(1) the density contrast field is Gaussian, the probability to find 
the density contrast $\delta$ (in a sphere with center in $x$ and radius $R$) 
larger than a fixed value $\delta_c$ (White\cite{15}) is given by the formula:
$$
f(R)=\large \int\limits_{\delta_c}^{\infty}\frac{1}{\sqrt{2\pi} 
\sigma(R)}\exp\left(-\frac{\delta^{2}}{2\sigma(R)^{2}}\right)d\delta;
$$
(2) the mass distribution of gravitationally bounded objects 
(``halos'') is the same as the distribution of density peaks.
In other words, the collapse of matter is spherical. 

To calculate $\delta_c$ in standard P\&S formalism, the linear overdensity of
object $\frac{\delta\rho}{\rho}$ in the uniform background at the collapse 
moment is considered. In the spatially  flat matter-dominated Universe 
(Gunn \& Gott\cite{3}) $\delta_c=1.686$, cases of spatially flat $\Lambda$CDM 
($\Omega_{tot} \equiv \Omega_m+\Omega_{\Lambda}=1$) and open
cosmological models ($\Omega_{tot}<1$) were studied in (Madau et al\cite{8}, 
Eke et al\cite{2}, Lacey \& Cole\cite{4}, Lahav et al\cite{6}, Lokas \& 
Hoffman\cite{5}). 

In standard form P\&S formalism allows to calculate 
the averaged mass function in the homogeneous Universe. This standard 
formalism does not take into account the existence of inhomogeneities (such as
voids and superclusters), where matter densities ($\rho_{v}$ and $\rho_{scl}$) 
are different from the background value ($\rho_0$). 
Now let us modify P\&S formalism and describe the influence of inhomogeneous
background related with existence of voids and superclusters. 

To analyze this influence we need: 
(i) to separate LSS background and density perturbations 
related with gravitationally bounded halos;
(ii) to calculate the growth rates of density perturbations 
in voids and superclusters separately.

The first step is rather simple one. Let us decompose the density contrast 
as follows: $\delta(R)=\delta(L)+\delta(R,L),$ where $\delta(L)$ is density 
contrast value related with void or supercluster with linear scale $L$, 
$\delta(L,R)$ is density contrast related with gravitationally bounded halos 
with linear scale $R$ in void or supercluster. 

The dispersion of the halo density contrast is:
\begin{equation}
\sigma(R,L)^2=\frac{1}{2\pi^2}\int\limits_{0}^{\infty} P(k) |W(kR)-W(kL)|^2 k^2 dk ,
\end{equation}
Thus, the mass function of halo in voids and superclusters as follows:
\begin{equation}
n_{L}(M) = \sqrt{\frac{2}{\pi}}\frac{\rho_{l}\delta_{c,L}}{M}
 \frac{1}{\sigma(R,L)^{2}} \left| \frac{\partial \sigma(R,L)}{\partial M} 
\right| \exp\left(-\frac{\delta^2_{c,L}}{2\sigma(R,L)^2}\right), 
\end{equation}
here and further the subscriptions $''L''$ means a size of local region. It is
clear from the Fig.1 that if $L\rightarrow\infty$ we deal with standard P\&S 
curve. 
\begin{figure}[t]
\epsfysize=0.4\hsize 
\epsfxsize=7.5cm       
\centerline{\epsfbox{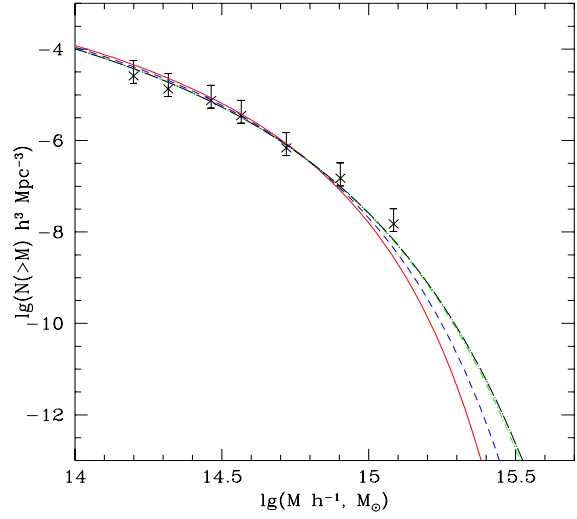}}
\caption{$N(>M)$ versus $L$. Solid line corresponds to $L=30$ Mpc,
dashed line to $L=50$ Mpc, dot-dashed line to $L=100$ Mpc, 
dot-dot-dashed line corresponds to the standard P\&S curve. 
$\Omega_\Lambda=0$, $\delta_c=1.686$.}
\end{figure}

To calculate $\delta_{c,L}$ in void or supercluster we analyze the collapse 
of gravitationally bounded object in under- and over- density regions. 

\section{Collapse in the inhomogeneous Universe}
\subsection{The case of matter-dominated Universe}

Radii of void and supercluster evolve as follows (Landau \& Lifshits \cite{16}):
\\
\noindent void: 
\begin{equation}
\frac{R}{R_{v}}=\frac{1}{2}({\rm ch}\;\eta-1),
\;\;\;
\frac{t}{t_{v}}=\frac{1}{\pi}({\rm sh}\;\eta-\eta), 
\end{equation}
\noindent supercluster:
\begin{equation}
\frac{R}{R_{scl}}=\frac{1}{2}(1-\cos\;\eta), 
\;\;\;
\frac{t}{t_{scl}}=\frac{1}{\pi}(\eta-\sin\;\eta), 
\end{equation}
where $R$ is radius of void or supercluster, $t$ and $\eta$ are physical and 
conformal time (physical time is different in voids and superclusters), 
$t_{scl}$, $t_v$, $R_{scl}$ and $R_v$ are constants. 

Using the methodology for calculation of density threshold in uniform 
background (White\cite{15}) we applied it for the case of inhomogeneous 
background. Assuming $\eta\rightarrow 0$ and decomposing (3) and (4) 
we derive for void/supercluster:
$$
\frac{R}{R_{v,scl}}=\frac{t^{\frac{2}{3}}}{R_{v,scl}}\left\{1 \pm 
\frac{1}{20}\left(\frac{6\pi t}{t_{v,scl}}\right)^{\frac{2}{3}} \right\}.
$$
As the density contrast related with halo in void/supercluster is as 
follows: 
$$
\delta_{L}\equiv\frac{|\rho-\rho_{L}|}{\rho_{L}}
=\frac{R^3_{L}}{R^3_{halo}}-1,
$$
where $\rho$ is matter density in halo, $R_{halo}$ is a radius of halo,
$L$ denots the void or supercluster.  
So, the critical density contrast related with halo in discussed regions 
are as follows:
\begin{equation}
\delta_{v,scl}=1.686 \left\{1 \pm \left(\frac{t}{t_{v,scl}}\right)^{\frac{2}{3}}\right\},
\end{equation}
where $\frac{t}{t_{v,scl}}$ is a function of total density 
$\Omega_{tot,L}=\Omega_{m,L}+\Omega_{m}$ in void or supercluster. 

To relate $\delta_{c,L}$ and $\Omega_{tot,L}$ we can use Friedmann equation 
$3H^2=4\pi G\rho$, where all quantities have to be calculated in local region:
$H_L=\frac{\dot R}{R}$, $\rho = \rho_m + \rho_{curv}$ where the first term 
is a matter, the second one is curvature.        

So, in the superclusters ($\Omega_{tot}>1$) we can rewrite the Friedman 
equation as follows:
$$
3H^{2}=\frac{3\pi^{2}t^2}{t_{scl}^{2}}\left[\frac{1}{(1-\cos\eta)^{3}}+
\frac{\cos\eta}{(1-\cos\eta)^{3}}\right],
$$
where the first term is a averaged density in the supercluster.
Finally, $\Omega_t$ in supercluster as follows:
\begin{equation}
\Omega_{tot}^{scl} \equiv\frac{\rho_{scl}}{\rho_{cr}}=
\frac{9\pi^2}{4}\frac{t^2}{t_{scl}^2}\left(\frac{1}{1-\cos\eta}\right)^3. 
\end{equation}
For void ($\Omega_{tot}<1$) similar calculations give rise to:
\begin{equation}
\Omega_{tot}^{v}=
\frac{9\pi^2}{4}\frac{t^2}{t_v^2}\left(\frac{1}{1-{\rm ch}\;\eta}\right)^3,
\end{equation}
where $(\frac{t}{t_{v,scl}})$ and $\eta$ are related by (6) and (7).
Taken together equations (3)-(7) allow to relate $\delta_{c,L}$ and 
$\Omega_{tot,L}$ in voids and superclusters for cosmological models without 
$\Lambda$-term (see dotted line on the Fig.2). 

\begin{figure}[t]
\epsfysize=0.4\hsize 
\epsfxsize=7.5cm       
\centerline{\epsfbox{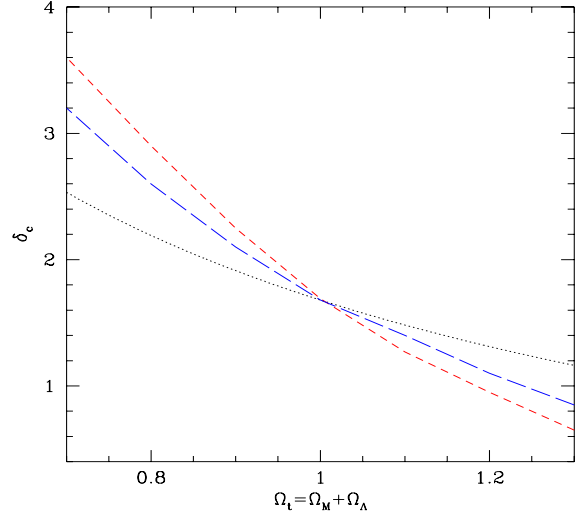}}
\caption{ The density contrast threshold as a function of $\Omega_{tot,L}$ in a
superclusters and voids. Dashed line corresponds to $\Omega_\Lambda$=0.6,
long dashed -- to $\Omega_\Lambda=0.3$, dotted curve -- to $\Omega_\Lambda=0$.}
\end{figure}

\subsection{The case of non-zero $\Lambda$-term} 

To calculate $\delta_{c,L}$ in the case we use the approach proposed by Lahav 
et al. (Lahav et al\cite{6}) and developed by others (Lacey \& Cole\cite{4}, 
Eke et al\cite{2}, Lokas \& Hoffman\cite{5}). Our calculations 
are mainly based on the last article (we refer it as LH00). 
To apply LH00 formalism we need to modify its basement. The first change is
 related to Hubble constant calculation. In LH00 $H$ does not depend on 
$\Omega_{tot}$ and is equal to observable value ($65\; km\;sec^{-1}Mpc^{-1}$).
In our case $H \equiv H_L$ is a function of local regions.
The second modification of LH00 is 
the introduction of the above-mentioned $L$, which is analogous to the
scale factor in voids or superclusters.

To calculate $H_{L}$ we solve Friedman equation: 
\begin{equation}
\left(\frac{H_{L}}{H_0}\right)^2=\frac{\Omega_{m,L}}{L^3}+
\Omega_{\Lambda} \pm \frac{\Omega_{m,L}}{L^2}, 
\end{equation}
$H_0$ is the current value of Hubble constant in the homogeneous Universe, 
$\Omega_{m,L}$ is matter density in void or supercluster.

To relate $L$ with time we need a similar formula for the evolution of Hubble parameter 
in the Universe:
\begin{equation}
\left(\frac{H}{H_0}\right)^2=\frac{\Omega_m}{a^3}+\Omega_{\Lambda}. 
\end{equation}

Assuming that physical time interval is the same in the different regions of 
the Universe it can be written as follows: $dt = ad\eta=Ld\tau,$ where $a$ is 
scale factor in the total Universe. Thus, we can rewrite last equation in 
more convenient form:
\begin{equation}
\int^\beta_0 \frac{d\beta}{\sqrt{(\frac{1}{\beta}+\beta^2 \pm c)}}
= \int^\alpha_0 \frac{d\alpha}{\sqrt{(\frac{1}{\alpha}+\alpha^2)}}, 
\end{equation}
where 
$\alpha \equiv \left(\frac{\Omega_{m,L}}{\Omega_{\Lambda}} 
\right)^{-\frac{1}{3}} a$,
$\beta \equiv \left( \frac{\Omega_{m,L}}{\Omega_{\Lambda}}
\right)^{-\frac{1}{3}} L$,
$c=\frac{(\Omega_{m,L}/\Omega_{\Lambda})^{\frac{1}{3}}}{a}$. The subscription 
$''0''$ means that the consequent quantity is taken now ($a_0=1$). Found 
$H_{L}$ and $L$ have to be substitute into eq.(19) of LH00. The result of 
numerical calculations is presented on Fig.3 (dashed and long dashed lines).

\section{Mass function in voids and superclusters}

Assuming results obtained in Section 2 and 3 we can calculate the new mass 
functions in void and supercluster separately. They are designed on Fig.3 
for cosmological model with $\Omega_m^{scl}=0.7$, 
$\Omega_m^{v}=0.1$, $\Omega_\Lambda=0.6$. All curves normalized 
by $\sigma_8$ (Eke et al. 1998). 

\begin{figure}[t]
\epsfysize=0.4\hsize 
\epsfxsize=7.5cm       
\centerline{\epsfbox{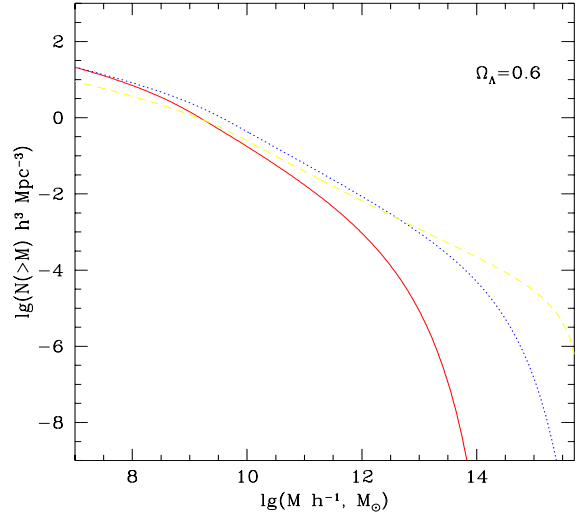}}
\caption
{
The galaxy cluster mass functions for the Universe regions with different 
$\Omega_{t,L}$. The dashed line corresponds to $\Omega_{t,L}=1.3$ 
(supercluster), the dotted line -- to $\Omega_{t,L}=1.0$ (flat homogeneous 
background), the solid line corresponds $\Omega_{t,L}=0.7$ (void). 
$\Omega_{\Lambda}=0.6$.
}
\end{figure}

It is clear from Fig.3 that the difference between mass functions 
calculated for slightly varying $\Omega_t$ is significant for all considered 
masses. In this case the ratio between mass functions in supercluters and 
voids can be approximated by exponential law:
$$
\frac{N_{scl}(>M)}{N_v(>M)}=20.1 \exp\left(\frac{M}{9.4\times10^{12}h^{-1} 
M_\odot}\right).
$$      
This effect increases in models with larger $\Omega_\Lambda$.

\section{Discussion and Conclusions}
To provide the correct calculation $N_L(>M)$ in local region 
we take into account the dependence of the threshold 
density contrast ($\delta_{c,L}$) on the mean matter density in considered 
regions ($\Omega_{m,L}$). 
We find with $10\%$ accuracy the following approximation for $\delta_{c,L}$:
$$
\delta_{c,L}=1.686-0.05\Omega_\Lambda-2.42\Delta(1+1.78\Omega_\Lambda)+
1.7\Delta^2(1+3.1\Omega_\Lambda),
$$
where $\Delta=\Omega_{m,L}-\Omega_m$. 

The significant difference between mass function in void and supercluster is 
found. We suggest an analytical approximation for the differential mass 
function in voids and superclusters for the spatially flat 
Universe with $\Omega_\Lambda=0.6$ (the accuracy is less than $10\%$:
$$
n(M)=3\cdot10^{-5}\cdot\left(\frac{M}{M_0}\right)^{-1.8}\exp\left(
-\frac{M}{M_0}\right),
$$   
where the parameter $M_0$ depends on $\Omega_{m,L}$:
$$
M_0=[(1+21.8\Omega_{m,L})^2+0.4]\cdot 2.5 \cdot 10^{12}.
$$

\section*{Acknowledgments}
This work was supported by Russian Foundation for Basic Research (01-02-16274).
Authors thank the Organizing Committee for financial support.


\end{document}